\documentclass[aps,prb,floatfix,amsmath,amssymb,preprint,showpacs]{revtex4-1}
\usepackage{amsmath, amssymb, graphics, mathrsfs, bm, stackrel,bbold}
\usepackage{subfigure}
\usepackage[usenames,dvipsnames]{xcolor}
\definecolor{Highlight}{rgb}{1,1,0.75}

\allowdisplaybreaks[1]
\usepackage{graphicx}
\usepackage[bookmarks={false}, pdfauthor={Rudro Rana Biswas}, pdftitle={Title}]{hyperref}
\hypersetup{colorlinks=true, linkcolor=BrickRed, citecolor=Violet, filecolor=OliveGreen, urlcolor=RoyalBlue, filebordercolor={.8 .8 1}, urlbordercolor={.8 .8 0}}
\usepackage[all]{hypcap}

\newcommand\ba{\begin{array}}
\newcommand\ea{\end{array}}
\newcommand\nn{\nonumber}

\newcommand\ri{\right}
\renewcommand\le{\left}

\newcommand{\feyn}[1]{#1\kern-0.45em/}

\renewcommand\a{\alpha}

\newcommand\mbA{\mbs{A}}
\renewcommand\b{\beta}

\renewcommand\c{\psi}

\renewcommand\d{\delta}

\newcommand\f{\phi}

\newcommand\G{\Gamma}

\renewcommand\k{\kappa}

\newcommand\n{\nu}

\newcommand\p{\pi}
\newcommand\mbp{\mbs{p}}

\newcommand\rr{\rho}

\newcommand\mbr{\mbs{r}}

\renewcommand\th{\theta}

\newcommand\mbx{\mbs{x}}

\newcommand\mc{\mathcal}
\newcommand\mb{\mathbb}
\newcommand\mbs{\boldsymbol}

\begin{document}
\title{Fractional charge and inter-Landau level states at points of singular curvature}
\author{Rudro R. Biswas}
\email{rrbiswas@anl.gov}
\affiliation{Materials Science Division, Argonne National Laboratory, Argonne, Illinois 60439, USA}
\author{Dam T. Son}
\email{dtson@uchicago.edu}
\affiliation{Kadanoff Center for Theoretical Physics, University of Chicago, Chicago, Illinois 60637, USA}

\begin{abstract}
The quest for universal signatures of topological phases is fundamentally important since these properties are robust to variations in system-specific details. Here we present general results for the response of quantum Hall states to points of singular curvature in real space. Such topological singularities may be realized, for instance, at the vertices of a cube, the apex of a cone, etc. We find, using continuum analytical methods, that the point of curvature binds an excess fractional charge. In addition, sequences of states split away, energetically, from the degenerate bulk Landau levels. Importantly, these inter-Landau level states are bound to the topological singularity and have energies that are \emph{universal} functions of bulk parameters and the curvature. Remarkably, our exact diagonalization of lattice tight-binding models on closed manifolds shows that these results continue to hold even when lattice effects are significant, where the applicability of continuum techniques could not have been justified a priori. Moreover, we propose how these states may be readily experimentally actualized. An immediate technological implication of these results is that these inter-Landau level states, being as they are \emph{both} energetically and spatially isolated quantum states, are promising candidates for constructing qubits for quantum computation.
\end{abstract}
\maketitle

\emph{Introduction:} Quantum Hall (QH) states \cite{2002-yoshioka-zl} were amongst the first known examples of topologically nontrivial quantum states, a rapidly expanding category that now includes other exotica, such as Chern and topological insulators\cite{2011-hasan-uq,2011-qi-uq}, topological superconductors\cite{2011-hasan-uq,2011-qi-uq} and spin liquids\cite{2010-balents-uq}. The distinguishing feature of QH states is that they possess quantized values of Hall conductance, which are rational multiples of the conductance quantum, $e^{2}/h$, a combination of fundamental physical constants. This characteristic property results from the interplay between the degeneracy of QH states and their topological response to phase twists across their boundaries\cite{1981-laughlin-yq,1985-niu-nr}.

An additional topological feature of QH states, less well-known and appreciated several years after the first, is that they possess a universal coupling to the geometry of the two-dimensional real space manifold where the electrons reside\cite{1992-wen-fk}. This novel coupling causes a real space curvature field, $K(\mbx)$, to induce an excess density,
\begin{align}\label{eq-wenzeedensity}
\d\rr(\mbx) \equiv \rr(\mbx) - \frac{\n}{2\p} = \frac{\k}{2\p}K(\mbx).
\end{align}
Here we set the magnetic length $\ell = \sqrt{\hbar/eB}$ to unity; $\n$, the filling fraction, is equal to one for an isolated full Landau level (LL). The value of the `gravitational' coupling constant, $\k$, is a characteristic of the QH state, and is stable to small perturbations to the Hamiltonian\cite{1992-wen-fk}. LLs with indices $n=0,1,\ldots$, are characterized by corresponding values\cite{1992-wen-fk,1994-iengo-fk} of the coupling constant, $\k_{n} = n+1/2$. Non-intuitive and interesting new behaviors resulting from this gravitational coupling are being increasingly appreciated: it has recently been shown to be the source of anomalous viscosity\cite{2009-read-fk}, and arises in the non-uniform electrical response of QH states\cite{2012-hoyos-fk, 2012-bradlyn-fk, 2013-biswas-wq,2014-can-qy}.

\begin{figure}[t]
\begin{center}
\subfigure[]
{\resizebox{6cm}{!}{\includegraphics[trim=1cm 0.5cm 0.5cm 0.5cm, clip=true]{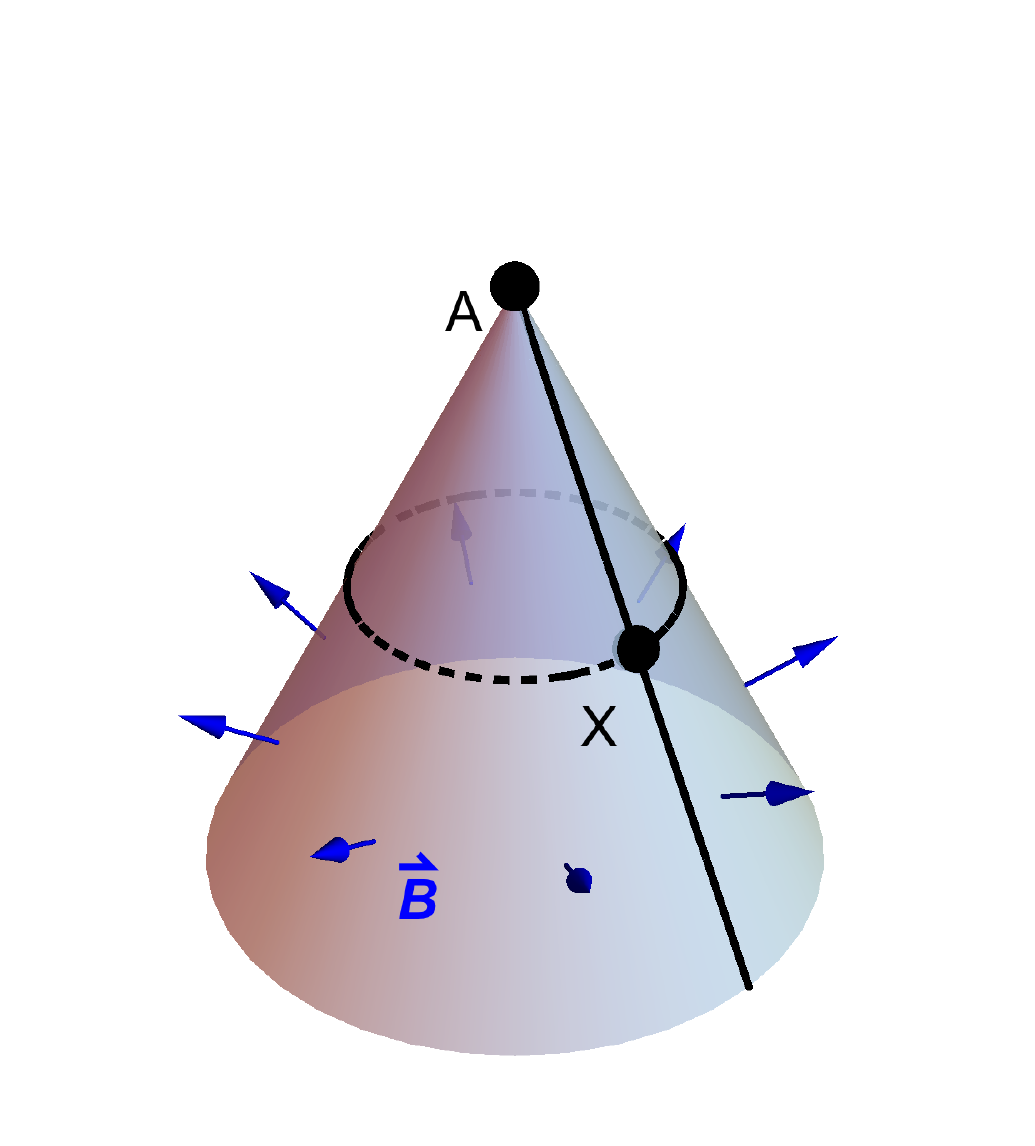}}
\label{fig-cone}}
\subfigure[]
{\resizebox{6cm}{!}{\includegraphics{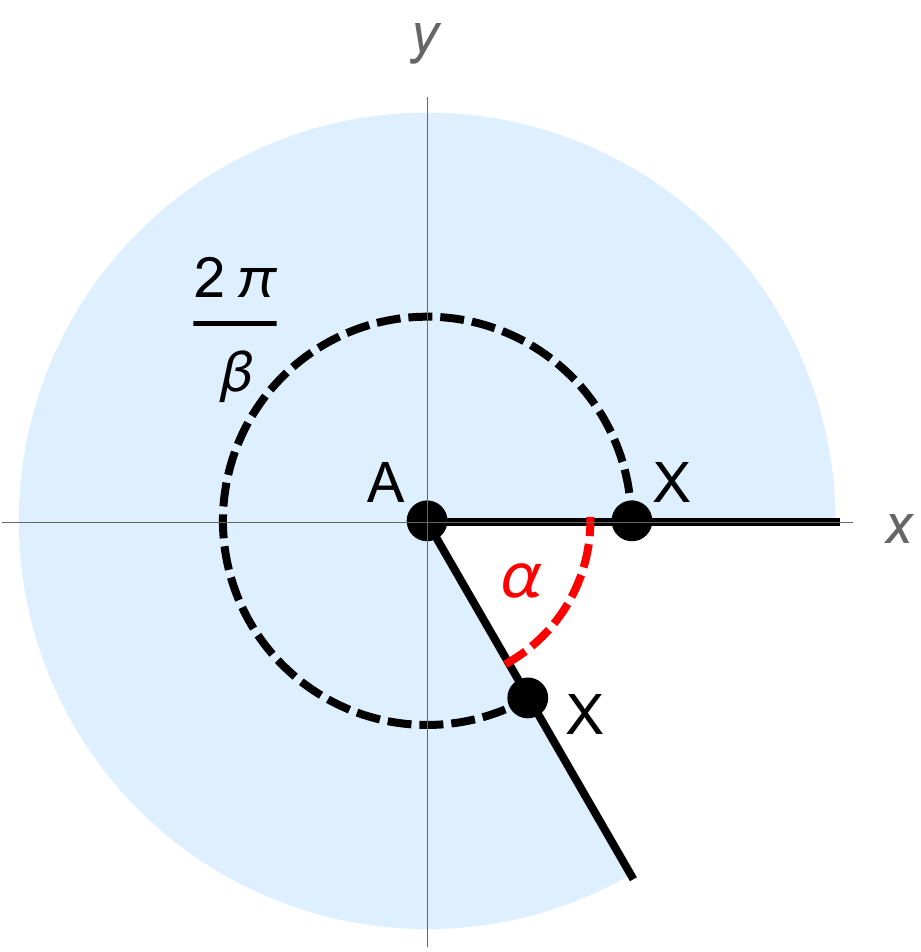}}
\label{fig-coneflattened}}
\caption{(Color online) Conical geometry being used in this paper. \subref{fig-cone} is a three-dimensional view of the conical surface, with the uniform magnetic field $\vec{B}$ applied perpendicular to it. \subref{fig-coneflattened} is a flattened view of the cone, which involves removing a wedge from the plane, with the apical angle of the wedge being equal to the deficit angle of the cone, $\a$.}
\label{fig-conesetup}
\end{center}
\end{figure}

\begin{figure}[t]
\begin{center}
\subfigure[]
{\resizebox{6cm}{!}{\includegraphics{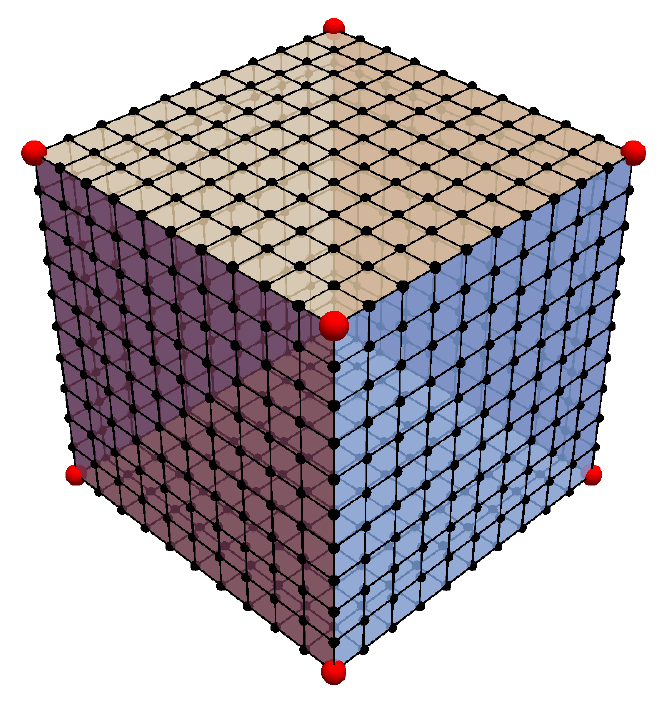}}
\label{fig-cube}}
\subfigure[]
{\resizebox{7cm}{!}{\includegraphics{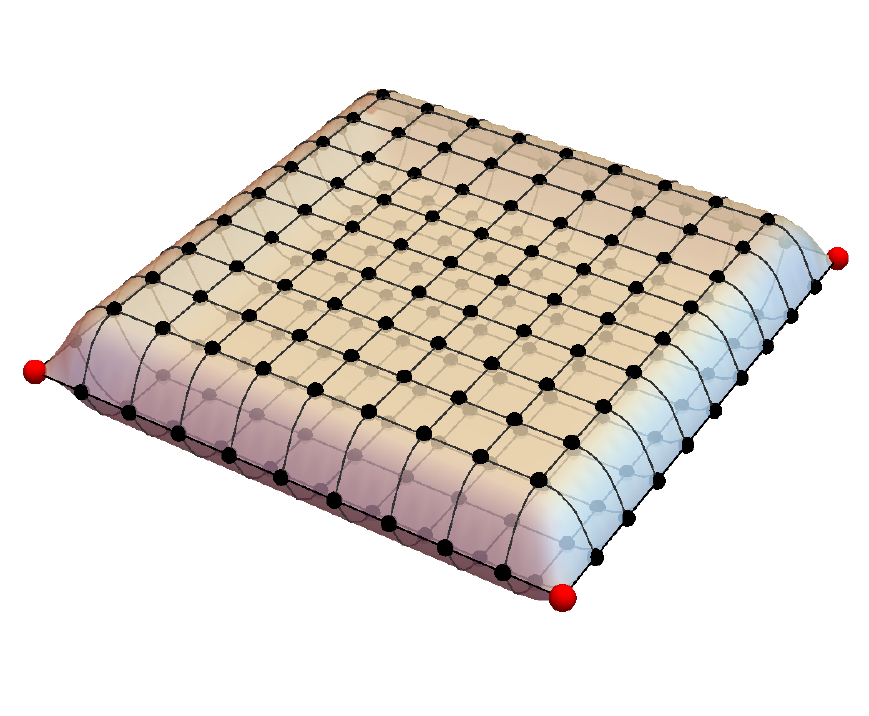}}
\label{fig-biplane}}
\caption{(Color online) Cube and biplane structures used for exact diagonalization studies. Large red sites at the vertices are locations of the corresponding disclinations, with deficit angles $\a = \p/2$ and $\p$ respectively. We have studied nearest neighbor tight binding Hofstadter models on each structure.}
\label{fig-polyhedra}
\end{center}
\end{figure}

Actualization of this `gravitational' response in a real physical system requires being able to induce real space curvature on a lattice. Doing so will typically introduce defects and strains in the material, masking the topological effects predicted by the Chern-Simons theory\cite{1992-wen-fk}, and introducing non-universal corrections. However, by confining the required lattice topological defects to a point in real space (while allowing for slight bond-bending away from it), curvature can be introduced without inducing extensive local modifications to the energetics. Thus, a convenient way to probe the topological response of QH states to real space curvature is to create a conical structure, Fig.~\ref{fig-cone}; its tip size needs to be smaller than both the magnetic length and the characteristic kinetic wavelength of the charge carriers. An example that meets these requirements, even in the presence of high magnetic fields (of the order of a flux quantum per lattice plaquette), is a vertex with exactly one disclination. Such structures are found, for example, at any corners of a cube or biplane, Fig.~\ref{fig-polyhedra}.

Motivated by these considerations, we consider topological aspects of the response of integer QH states to a point of singular curvature in real space, a physically relevant scenario not considered in previous field theoretical analyses. Much literature exists, however, for effects of disclinations in graphene\cite{2004-lammert-yu,2012-bueno-rm,2013-ruegg-xy}. The point of singular curvature is equivalent to the apex of a cone in real space, with a deficit angle $\a$, and a corresponding singularity in the curvature field, $K(\mbx)$, localized at the apex: $\iint_{\text{apex}} K(\mbx) dA = \a$, $K=0$ elsewhere. Evidently, in order to preserve the integrity of the lattice except at the singularity point, the deficit angles and consequently, the net curvature, can only assume discrete values.

\begin{figure}[t]
\begin{center}
\resizebox{9cm}{!}{\includegraphics{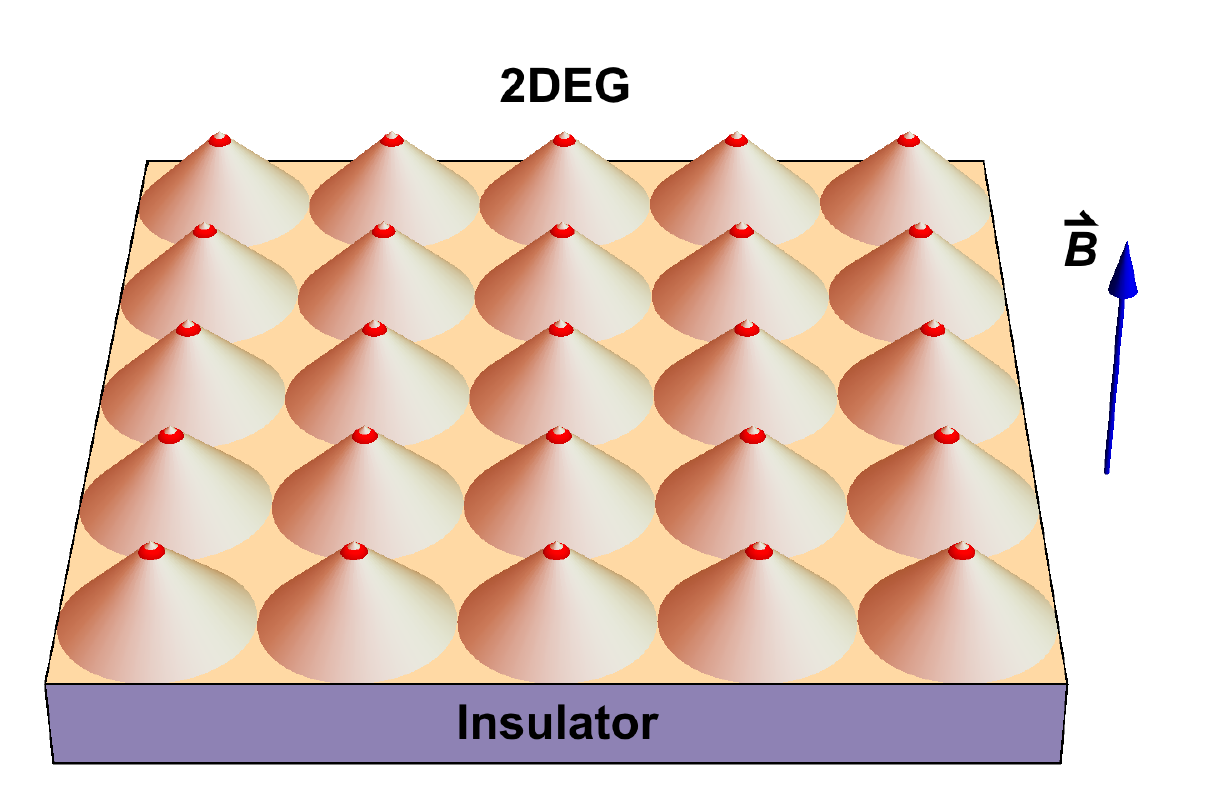}}
\caption{Suggested experimental setup for observing localized inter-LL states at points of singular spatial curvature (red tips). All features are large compared to the magnetic length.}
\label{fig-expsetup}
\end{center}
\end{figure}

In the continuum limit, i.e., when the magnetic and kinetic lengths are large compared to the lattice scale, we find that single particle states around such a singular point can be categorized into two types\cite{2012-bueno-rm}. Type I states, Eq.~\eqref{eq-eigenstatesI}, cause macroscopic degeneracy of the Landau level, providing a uniform density of $\rr_{0}=(2\p)^{-1}$ away from the singularity point. They are attracted by curvature and rearrange to give excess fractional charge near the singularity; its magnitude, Eq.~\eqref{eq-voidcharge}, is remarkably consistent with topological field theoretical analysis\cite{1992-wen-fk}. Type II states, Eq.~\eqref{eq-eigenstatesII}, are localized around the singularity. In the limit of vanishing curvature, they ensure that the Landau levels are completely degenerate, and have a fixed density, $\rr_{0}$, everywhere, including at the point of vanishing curvature. At non-zero curvature, type II states split away energetically from the bulk LL according to a universal prescription and (in general) move well into the inter-LL gap. The interplay of the charge responses of these two types of states to a small real space curvature, Eq.~\eqref{eq-totalcharge}, gives rise to the behavior predicted by Wen and Zee\cite{1992-wen-fk}. Remarkably, this picture holds quite well on the lattice, away from the continuum limit, as we find numerically. Creating artificial mesoscopic QH structures with such singularly curved tips, e.g. Fig.~\ref{fig-expsetup}, is feasible using currently available technologies; the novel behavior we outline above, especially the creation of inter-LL localized states with predictable and tunable energies, should thus be experimentally accessible by spectroscopic and other methods. These inter-LL states show promise for use in quantum computation, since they are \emph{both} energetically and spatially isolated quantum states.

\emph{Continuum calculation:} We consider non-interacting 2D charge carriers on the surface of a cone, with a uniform magnetic field, $\vec{B}$, perpendicular to the surface, as sketched in Fig.~\ref{fig-cone}. The cone surface can be mapped to a wedge subtending an angle $2\p/\b$ at the origin of a 2D plane, Fig.~\ref{fig-coneflattened}. There is uniform magnetic flux through its surface and particle wavefunctions (and all their derivatives) are matched across the edge $AX$. $\b$ is related to the deficit angle $\a$, and hence to the integrated curvature at the cone apex, 
\begin{align}\label{eq-totalcurvature}
\iint_{\text{cone tip}} K(\mbx) dA &= \a = 2\p \le(1-\frac{1}{\b}\ri).
\end{align}
In the continuum limit the long wavelength effective Hamiltonian is quadratic and takes the form
\begin{align}
\mc{H} &= \frac{(\mbp + e \mbA)^{2}}{2m}.
\end{align}
A convenient gauge choice for $\mbA$ is the symmetric gauge: $\mbA = (B r/2) \hat{\mbs{\th}}$. Here we use this gauge and a unit system where $m$, $e$, $\hbar$ and $B$ are all set to unity. Then the angular and radial variations of energy eigenstates may be separated as
\begin{align}
\c(r,\th) &= \f(r)e^{i \b m \th}, \; m\in\mb{Z}.
\end{align}
Quantization of the phase winding follows from the boundary matching condition $\c(r,2\p/\b) = \c(r,0)$, since the two loci of points $X$ in Fig.~\ref{fig-coneflattened} are, in fact, identical on the cone (Fig.~\ref{fig-cone}). The radial equation for $\f(r)$ may be solved using arguments similar to those used for finding simple harmonic oscillator wavefunctions\cite{1998-merzbacher-qe}, and in general has \emph{two types} of solutions\cite{2012-bueno-rm}.

\begin{figure}[t]
\begin{center}
\resizebox{10cm}{!}{\includegraphics{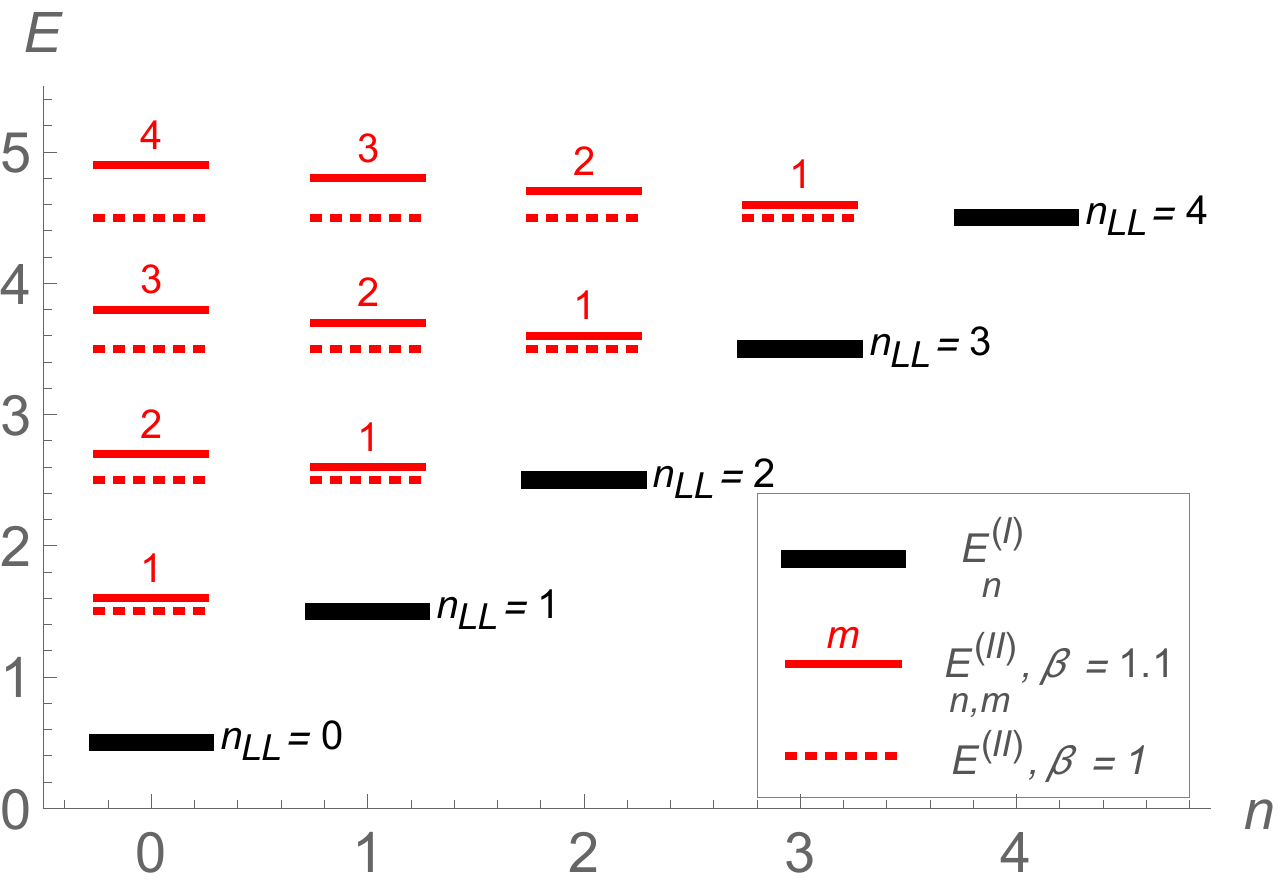}}
\caption{(Color online) Energy level structure for an almost-flat cone, $\b = 1.1$, compared to the planar case, $\b = 1$. Thick black bars denote type I states, Eq.~\eqref{eq-eigenstatesI}, that define macroscopically degenerate Landau levels whose energies, $E_{n}^{(I)}$, are independent of $\b$ (Eq.~\eqref{eq-eigenstatesI}). Red bars denote type II states (Eq.~\eqref{eq-eigenstatesII}), whose energies on the plane (dotted levels) are degenerate with the corresponding LLs, but on a cone generically occur inside the inter-LL gaps (unbroken levels). For such weak curvature and $n_{LL}<(\b-1)^{-1}$, we can still energetically identify the LL associated with the type II state.}
\label{fig-energiessmallbeta}
\end{center}
\end{figure}

\begin{figure}[t]
\begin{center}
\subfigure[]
{\resizebox{10cm}{!}{\includegraphics{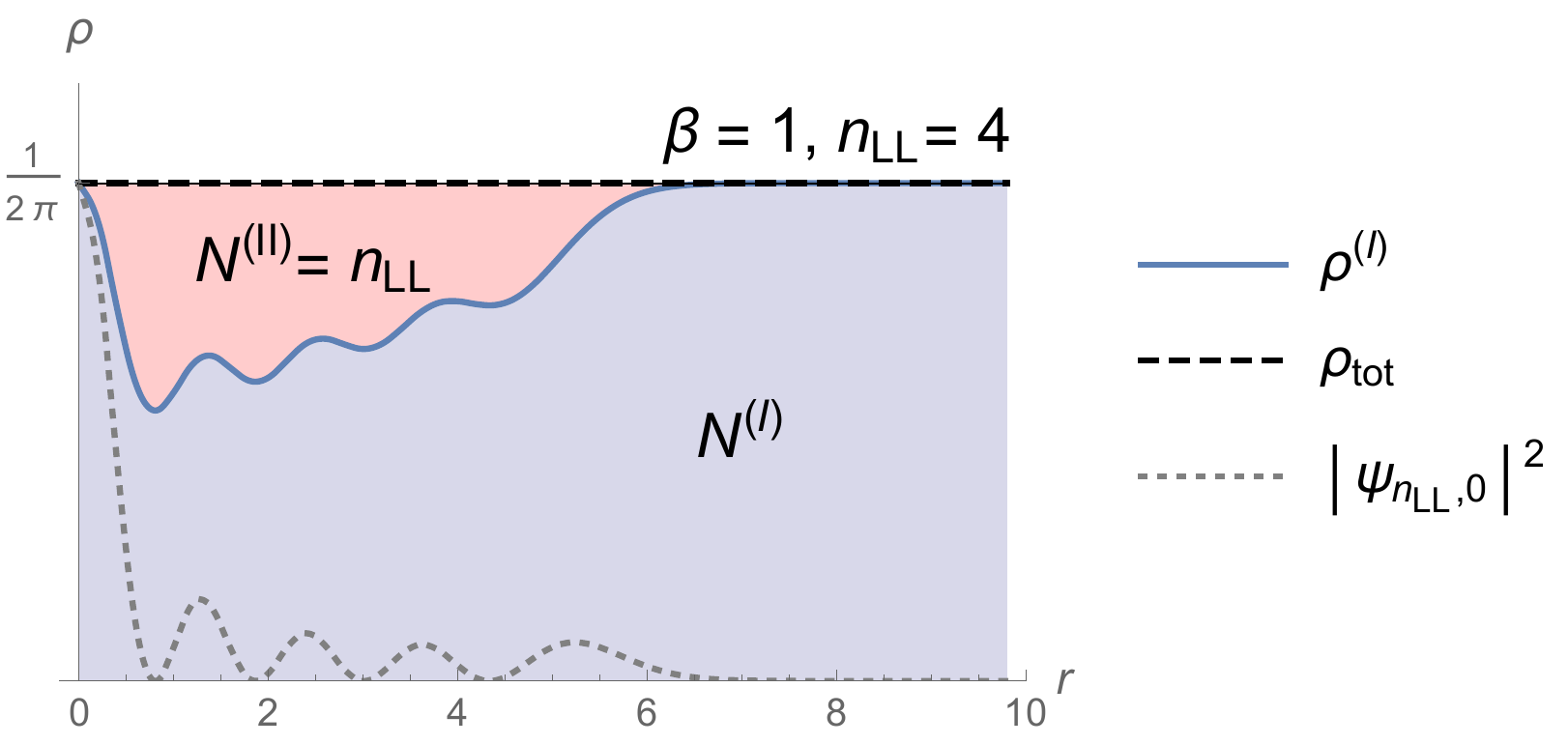}}
\label{fig-density1}}
\subfigure[]
{\resizebox{10cm}{!}{\includegraphics{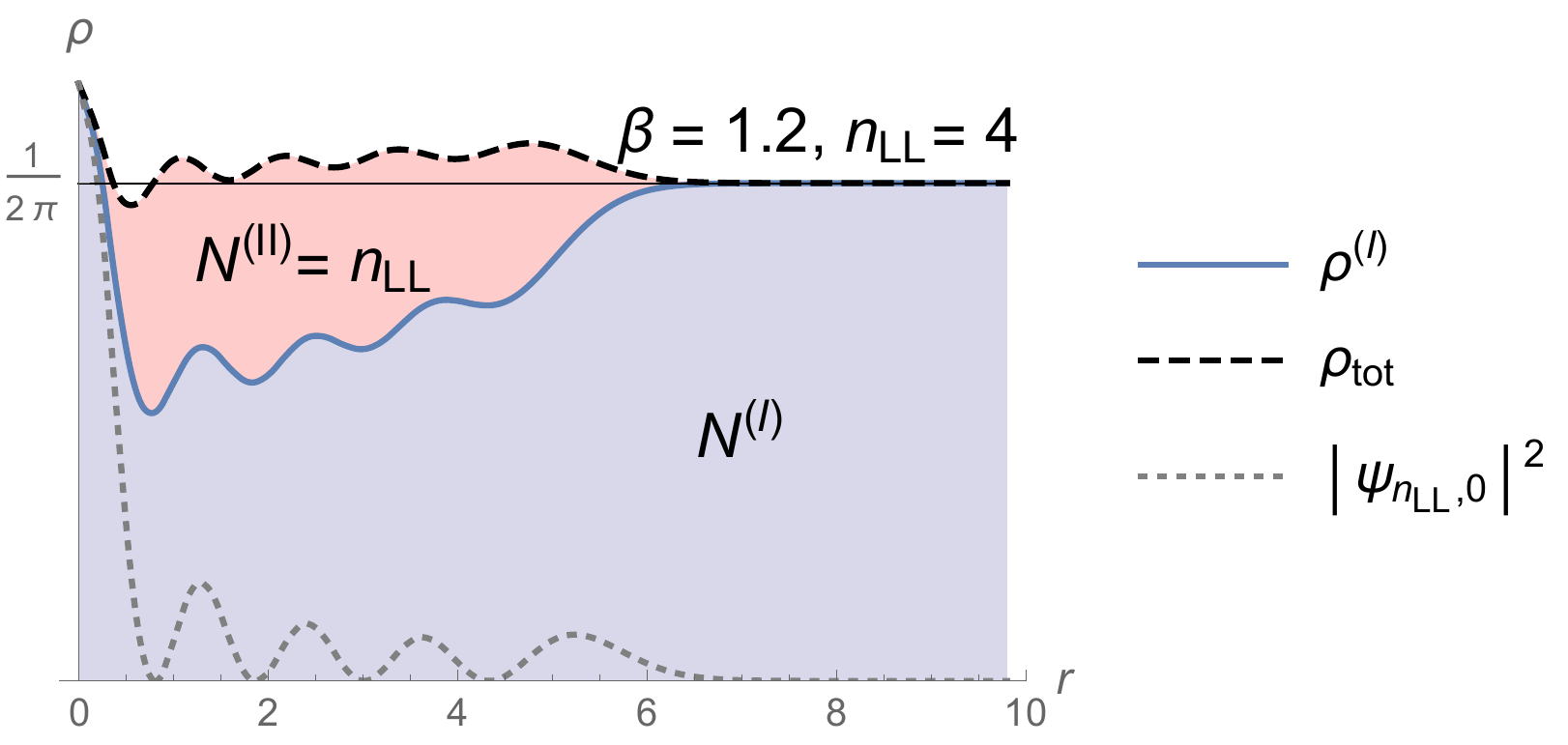}}
\label{fig-density2}}
\caption{(Color online) Radial density profiles of a filled LL on a cone, for the planar case \subref{fig-density1}, $\b=1$, and an almost planar case \subref{fig-density2}, $\b=1.2$. Type I states (shaded blue) leave a void near the cone apex at $r=0$, and account exactly for the uniform planar density value of $\rr_{0} = (2\p)^{-1}$ further away. The $n_{LL}$ type II states (shaded pink) fill the void near the cone apex. They add to give the density $\rr_{0}$ everywhere on the plane, and a little excess amount for $\b>1$, as evaluated in Eq.~\eqref{eq-totalcharge} and clearly evident in \subref{fig-density2}. Finally, only the $m=0$ state (gray, dashed) has a finite amplitude at the cone apex.}
\label{fig-densities}
\end{center}
\end{figure}

The `type I' energy eigenstate wavefunctions, expressed in terms of associated Laguerre polynomials, $L$, are
\begin{subequations}\label{eq-eigenstatesI}
\begin{align}
\c^{(I)}_{n,m}(r,\th) &= \mc{N}_{n,m} e^{- i \b m \th}e^{-r^{2}/4}r^{\b m} L_{n}^{\b m}(r^{2}/2),\\
E^{(I)}_{n} &= n +  \frac{1}{2}, \; n,m = 0,1, 2 \ldots,
\end{align}
\end{subequations}
where the normalization is $\mc{N}_{n,m}^{2} = \frac{\b n!}{2^{\b m +1}\p \G(n+\b m + 1)}$. Since their energies, $E^{(I)}_{n}$, are independent of $m$, these states cause the macroscopic degeneracy of the Landau levels. When all type I states in a LL with index $n_{LL} \equiv n$ are occupied, they give rise to a uniform density of $\rr_{0}=(2\p)^{-1}$ outside a radius (that itself grows as $\sqrt{n_{LL}}$), and a depleted region inside that radius (see Fig.~\ref{fig-densities}). The number deficit around the origin, when compared to the planar limit of a uniform density of $\rr_{0}$ everywhere, is
\begin{align}\label{eq-voidcharge}
\d N^{(I)}\le(n_{LL}\ri) &= \frac{1}{2} - \le(n_{LL} + \frac{1}{2}\ri)\frac{1}{\b}.
\end{align}
Here, $\d N^{(I)}\le(n_{LL}\ri) = \iint (\rr^{(I)}_{n_{LL}}(\mbr) - \rr_{0})dA$, with $\rr^{(I)}_{n}(\mbr) = \sum_{m=0}^{\infty}|\c^{(I)}_{n,m}(r,\th)|^{2}$. We have checked this result numerically for $n_{LL}\leq 10$, and analytically for $n_{LL}\leq 6$.

On a plane $\b = 1$ and the electron number deficit is exactly equal to $- n_{LL}$. Particle states in this void around the origin are accounted for by the type II eigenstates:
\begin{subequations}\label{eq-eigenstatesII}
\begin{align}
&\c^{(II)}_{n,m}(r,\th) = \mc{N}_{n,m} e^{ i \b m \th}e^{-r^{2}/4}r^{\b m} L_{n}^{\b m}(r^{2}/2),\\
&E^{(II)}_{n,m} = n + \b m + \frac{1}{2}, \; n= 0,1 \ldots, m = 1, 2,\ldots .
\end{align}
\end{subequations}

Compared with the type I solutions Eq.~\eqref{eq-eigenstatesI}, the azimuthal winding sense is reversed and the energies are no longer independent of $m$. On the plane, where $\b = 1$, both types of states are degenerate and the LL index for type II states is given by $n_{LL} = n+m$. Since $n\geq 0$ and $m \geq 1$, there are $N^{(II)}(n_{LL}) = n_{LL}$ type II states in the LL, accounting exactly for the number deficit $\d N^{(I)}\le(n_{LL}\ri) = -n_{LL}$ when only type I states are considered. In addition, incorporating these states makes the planar LL density uniform and and equal to $\rr_{0}=(2\p)^{-1}$ everywhere, Fig.~\ref{fig-density1}.

When the deficit angle is nonzero, these type II states, localized near the cone apex, move into the inter-LL gap. For small values of the curvature, $|\b - 1| \ll 1$, and for low enough LL index, $n_{LL} < 1/\le|\b - 1\ri|$, we can still identify the main LL associated with the inter-LL states, Fig.~\ref{fig-energiessmallbeta}. For this case, filling the LL gives rise to a curvature-induced excess particle number near the cone apex,
\begin{align}\label{eq-totalcharge}
&\d N(n_{LL}) = \d N^{(I)}(n_{LL}) + N^{(II)}(n_{LL}) \nn\\
&= \le(n_{LL} + \frac{1}{2}\ri)\le(1 - \frac{1}{\b}\ri) \equiv \frac{\k_{n_{LL}}}{2\p}\iint K(\mbx) dA.
\end{align}
We have used Eqs.~\eqref{eq-voidcharge} and \eqref{eq-totalcurvature}, and $\k_{n} = n+1/2$\cite{1992-wen-fk}. This is \emph{identical} to the topological result obtained by integrating the Wen-Zee formula\cite{1992-wen-fk}, Eq.~\eqref{eq-wenzeedensity}. An example elucidating how this charge excess arises is sketched in Fig.~\ref{fig-density2}.

Deviating further from the planar limit changes the energy level structure considerably, since type II states (red bars in Fig.~\ref{fig-energiessmallbeta}) now move far away from the LLs that they were originally degenerate with on the plane. We emphasize that this change in energies has a \emph{topological} origin, and arises independently of the bond-rearrangement that should also occur at the cone tip in a physical realization. Also, only the $\c_{n,0}$ states have finite amplitude at the cone apex, as can be readily seen from Eq.~\eqref{eq-eigenstatesI} and Fig.~\ref{fig-densities}. Thus, bond rearrangement at the apex will produce non-universal corrections to the energies of only the $\c_{n,0}$ states, one per LL.

\begin{figure}[t]
\begin{center}
\subfigure[]
{\resizebox{10cm}{!}{\includegraphics{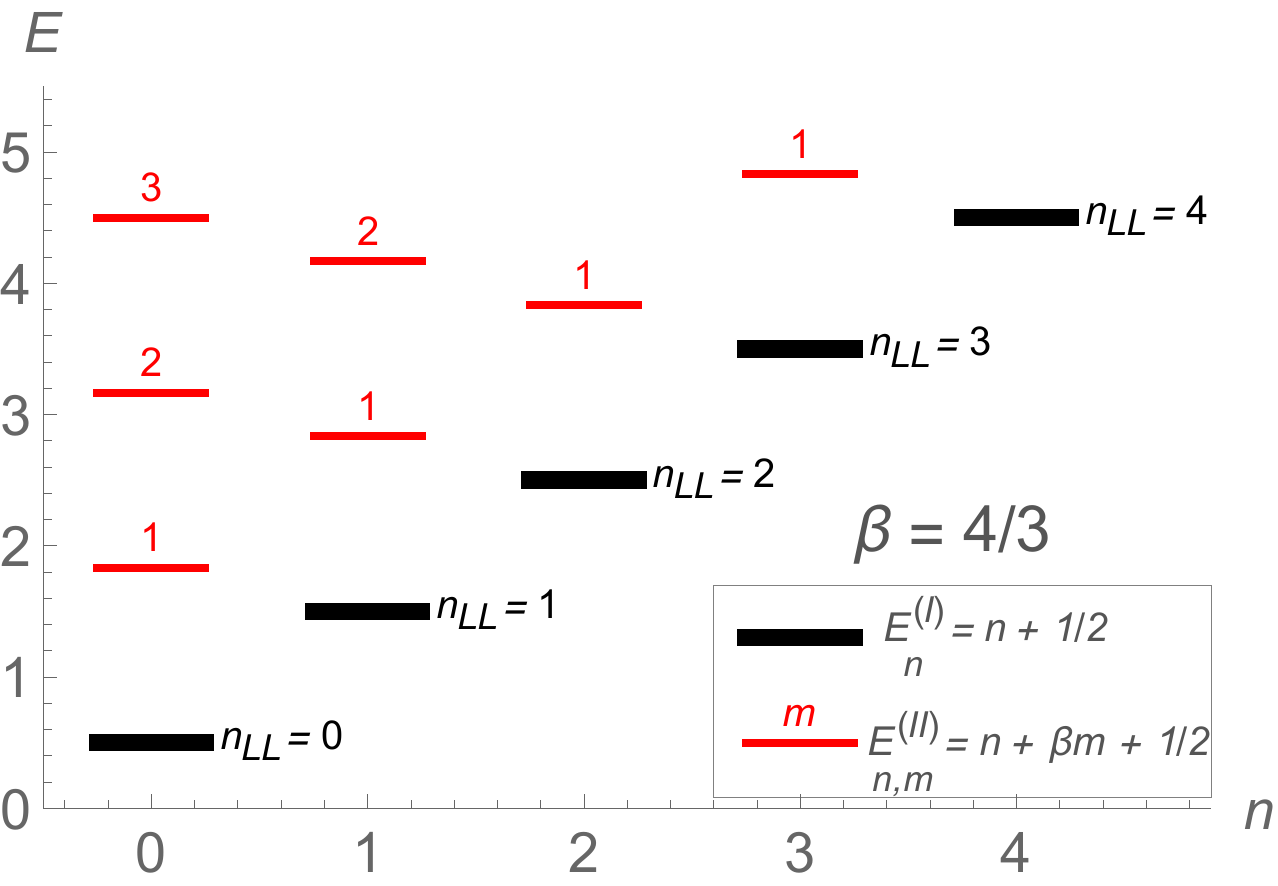}}
\label{fig-cubespectrumanalytic}}
\subfigure[]
{\resizebox{14cm}{!}{\includegraphics{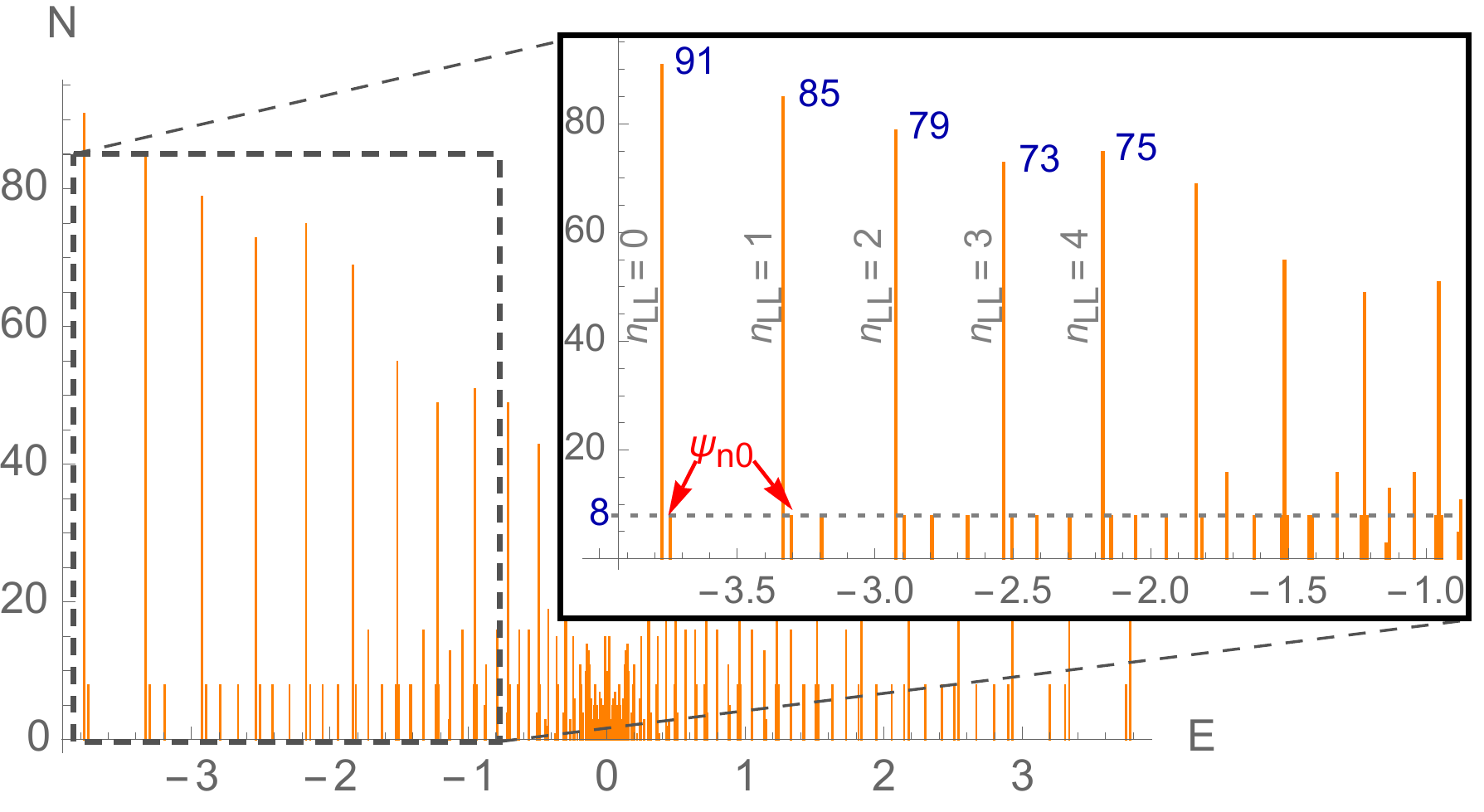}}
\label{fig-cubespectrumnumeric}}
\caption{(Color online) Comparison between \subref{fig-cubespectrumanalytic} analytic and \subref{fig-cubespectrumnumeric} exact diagonalized spectra on a cube surface. For \subref{fig-cubespectrumnumeric} the net magnetic flux is $N_{\f} = 98$ quanta and the cube side length is $L=20$. The magnetic length is thus $\ell = 2.9$, in lattice units. Type I states with LL index $n_{LL}$ have degeneracy $N^{(I)} = N_{\f} + 1 - 6n_{LL}$, also accounting for the $m=0$ states, some marked as $\c_{n0}$ here, whose energies are shifted by non-universal amounts. Type II states occur at $1/3$ and $2/3$ fractions of the gaps, as well as inside the LLs themselves.}
\label{fig-cubespectrum}
\end{center}
\end{figure}

\emph{Numerical calculation:} We have checked these statements via exact diagonalization calculations involving nearest-neighbor tight-binding Hofstadter models on a lattice. We use \emph{closed} manifolds, thus eliminating complications from edge states. Here we present two examples, using a square lattice. The square lattice can have two kinds of disclinations giving rise to positive curvature, with deficit angles $\a = \p/2$ or $\p$, i.e., $\b = 4/3$ or $2$. These can be used to construct macroscopic convex objects that are topologically equivalent to a sphere, and correspond to taking out a single quadrant, or an entire half of the planar lattice, respectively, and then folding the lattice appropriately around the corner. The regular objects that arise from these two types of disclinations are the cube, Fig.~\ref{fig-cube}, requiring eight disclinations of the first kind, and the `biplane', Fig.~\ref{fig-biplane}, having four vertices of the second kind. Since these are closed surfaces, an integer number of flux quanta, $N_{\f}$, are allowed through their surface. The solenoidal nature of the magnetic field is accounted for by using `Dirac strings', making numerical implementation of these models non-trivial\cite{2008-avishai-fk}.

\begin{figure}[b]
\begin{center}
\subfigure[]
{\resizebox{10cm}{!}{\includegraphics{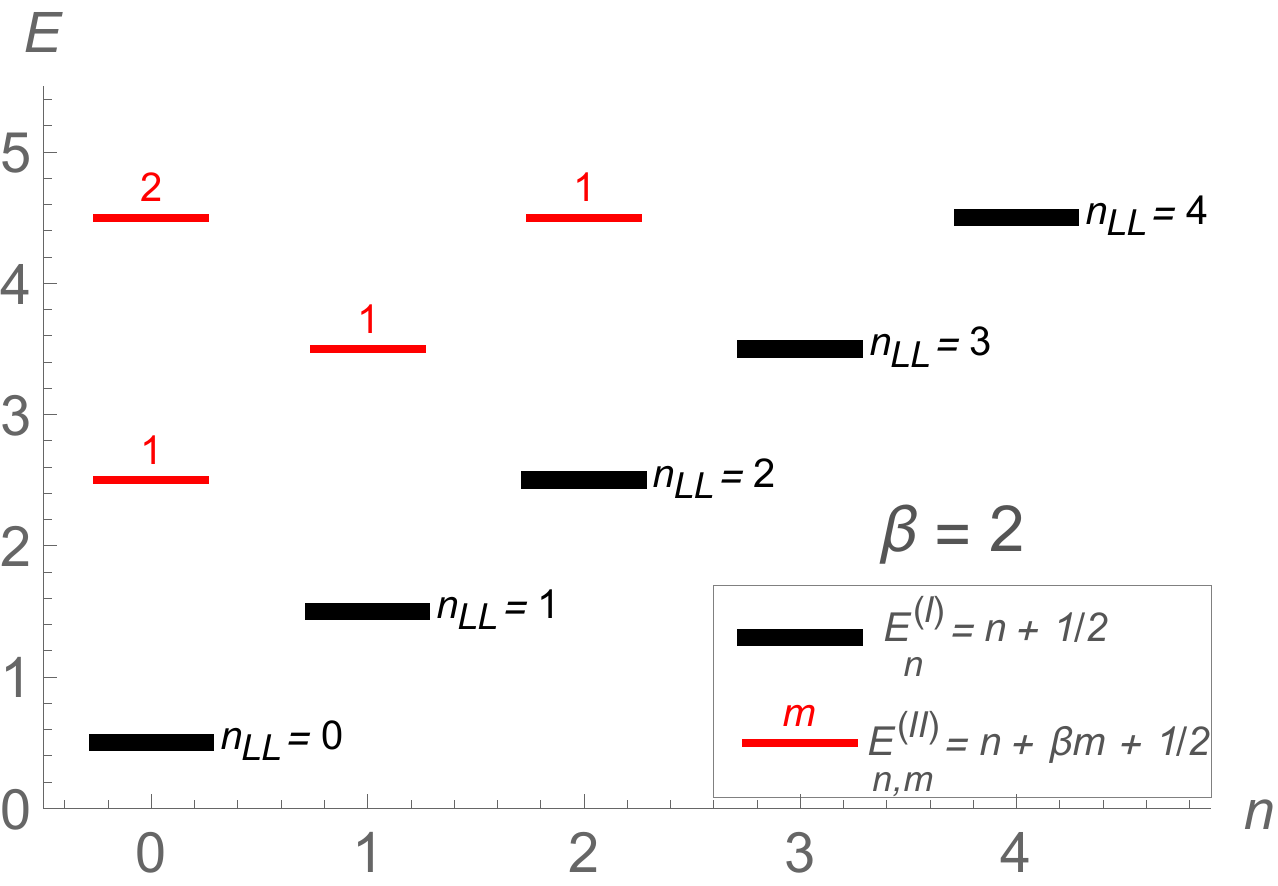}}
\label{fig-biplanespectrumanalytic}}
\subfigure[]
{\resizebox{14cm}{!}{\includegraphics{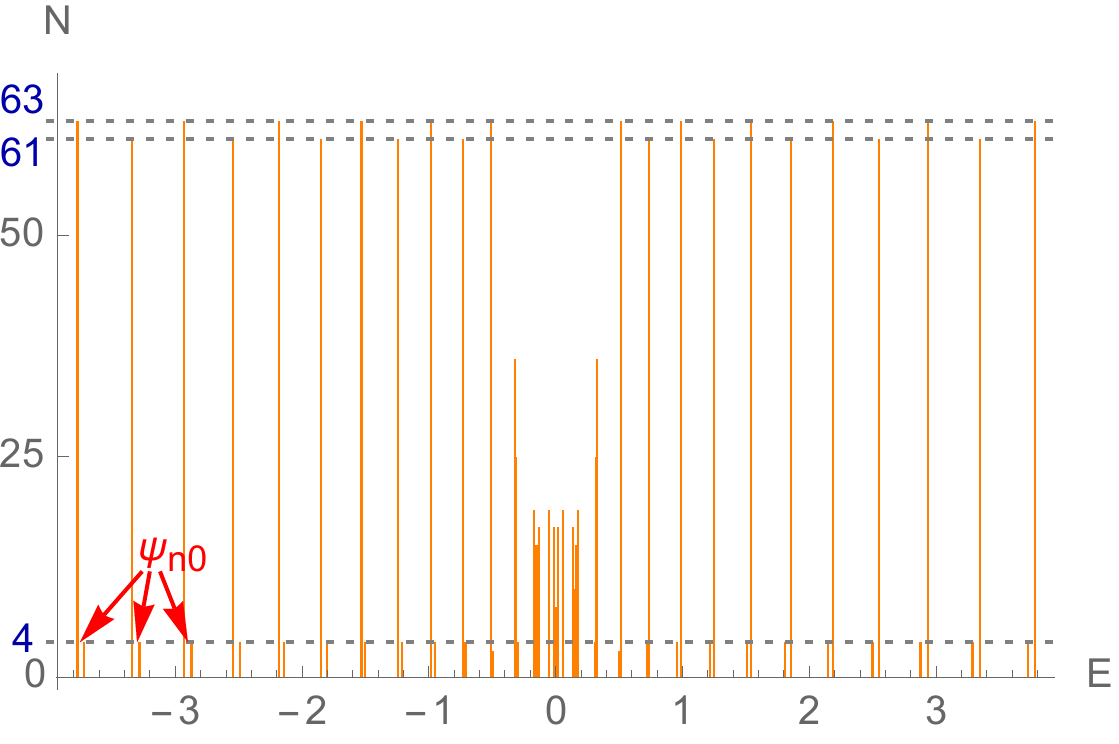}}
\label{fig-biplanespectrumnumeric}}
\caption{(Color online) Comparison between \subref{fig-biplanespectrumanalytic} analytic and \subref{fig-biplanespectrumnumeric} exact diagonalized spectra on a biplane surface. For \subref{fig-biplanespectrumnumeric}, $N_{\f} = 66$ quanta and biplane edge length $L=30$ ($\ell = 2.1$). Type I and II states are degenerate, except for non-universal energy corrections to the $m=0$ states (some marked as $\c_{n0}$), and the total LL population is simply $N_{\f} + (-1)^{n_{LL}}$.}
\label{fig-biplanespectrum}
\end{center}
\end{figure}

Our observations for the cube are presented in Fig.~\ref{fig-cubespectrum}. The Hofstadter spectrum is particle-hole symmetric, and $n_{LL}$ increases inwards in steps of $1$ from $n_{LL}=0$ at the outermost energies (Fig.~\ref{fig-cubespectrumnumeric} inset). Continuum analysis at any of the 8 vertices yields Fig.~\ref{fig-cubespectrumanalytic}. Using this with Eq.~\eqref{eq-totalcharge} at each vertex and noting that the uniform planar density of $\rr_{0}=(2\p)^{-1}$ corresponds to one particle per flux quantum, we expect, (i), type I states in a LL with index $n_{LL}$ contribute a total of $N^{(I)} = N_{\f} + 1 - 6n_{LL}$ states, and, (ii), type II states occur in multiplets of 8, at energies located at $1/3$ or $2/3$ of the inter-LL gap, or coincident with the bulk LL energies. As previously noted, since bond structure at each vertex is disrupted, the $m=0$ states have non-universal modifications to their energies, resulting in additional multiplets of 8 states per LL appearing in the inter-LL gaps (some marked as $\c_{n0}$ in Fig.~\ref{fig-cubespectrumnumeric}). These predictions are spectacularly borne out by the numerical results Fig.~\ref{fig-cubespectrumnumeric}, even when inter-LL gaps are non-uniform, as particle dispersion is no longer quadratic because of lattice effects. At higher fields (not shown), these levels spread out into bands but the counting pattern discussed above continues to hold.

Our observations for the biplane are shown in Fig.~\ref{fig-biplanespectrum}. For this case, $\b=2$ is an integer and the type II energy levels coincide with the main LL energies (Eq.~\eqref{eq-eigenstatesII}). The LL populations in this case is very simple: $N_{LL} = N_{\f} + (-1)^{n_{LL}}$. This is manifestly different from LLs on a sphere, where the degeneracy is $N_{\f} + 2n_{LL} + 1$\cite{1983-haldane-rt}. Keeping track of the quadruplets of $m=0$ states that receive non-universal energy corrections from the differing bond structure at the 4 corners, we see from Fig.~\ref{fig-biplanespectrumnumeric} that the numerical result indeed matches expectations from our analytic calculations. 

In conclusion, we have investigated the response of integer QH states to points in real space where curvature is singular, using both analytic and numerical techniques. Synthesizing such conical structures is an experimentally feasible method for investigating the gravitational response of QH states, till date  a purely theoretical enterprise. A schematic of such a setup is sketched in Fig.~\ref{fig-expsetup}, where the physics discussed in this work occurs locally near the cone tips (red dots). It is worth mentioning here that the curvature can also be made negative by creating saddle point structures, where our results will continue to hold. We have found that that, (i), excess fractional charge gets bound to the cone apex and, (ii), LL degeneracy is locally disrupted, leading to a finite number of states splitting away from the main LL near the apex.  These energy splittings are universal functions of curvature and cyclotron energy. These states, which are both spatially localized and energetically isolated by a tunable amount, are promising candidates for applications in quantum computation. Further, in analogy with how the shift\cite{1992-wen-fk} is used to characterize topological states with rotational invariance, we propose that the fractional part of the charge accumulated at disclinations may be used to characterize topological states on a lattice. Promising avenues for further research are  generalizing  these universal ideas to graphene and other Dirac materials, Chern insulators and fractional QH states, and also mapping out the interplay between orbital and Zeeman magnetic couplings.

\begin{acknowledgments}
We thank W. Kang, P. Wiegmann, L. Kadanoff and S. Iyer-Biswas for useful discussions. Work performed at Argonne National Laboratory (R.R.B.) is supported by the U. S. Department of Energy, Office of Science, Office of Basic Energy Sciences, under Contract No. DE-AC02-06CH11357. The work of D.T.S. is supported, in part, by DOE under grant No. DE-FG02-13ER41958, and by a Simon Investigator grant from the Simons Foundation.
\end{acknowledgments}


\begin{thebibliography}{19}%
\makeatletter
\providecommand \@ifxundefined [1]{%
 \@ifx{#1\undefined}
}%
\providecommand \@ifnum [1]{%
 \ifnum #1\expandafter \@firstoftwo
 \else \expandafter \@secondoftwo
 \fi
}%
\providecommand \@ifx [1]{%
 \ifx #1\expandafter \@firstoftwo
 \else \expandafter \@secondoftwo
 \fi
}%
\providecommand \natexlab [1]{#1}%
\providecommand \enquote  [1]{``#1''}%
\providecommand \bibnamefont  [1]{#1}%
\providecommand \bibfnamefont [1]{#1}%
\providecommand \citenamefont [1]{#1}%
\providecommand \href@noop [0]{\@secondoftwo}%
\providecommand \href [0]{\begingroup \@sanitize@url \@href}%
\providecommand \@href[1]{\@@startlink{#1}\@@href}%
\providecommand \@@href[1]{\endgroup#1\@@endlink}%
\providecommand \@sanitize@url [0]{\catcode `\\12\catcode `\$12\catcode
  `\&12\catcode `\#12\catcode `\^12\catcode `\_12\catcode `\%12\relax}%
\providecommand \@@startlink[1]{}%
\providecommand \@@endlink[0]{}%
\providecommand \url  [0]{\begingroup\@sanitize@url \@url }%
\providecommand \@url [1]{\endgroup\@href {#1}{\urlprefix }}%
\providecommand \urlprefix  [0]{URL }%
\providecommand \Eprint [0]{\href }%
\providecommand \doibase [0]{http://dx.doi.org/}%
\providecommand \selectlanguage [0]{\@gobble}%
\providecommand \bibinfo  [0]{\@secondoftwo}%
\providecommand \bibfield  [0]{\@secondoftwo}%
\providecommand \translation [1]{[#1]}%
\providecommand \BibitemOpen [0]{}%
\providecommand \bibitemStop [0]{}%
\providecommand \bibitemNoStop [0]{.\EOS\space}%
\providecommand \EOS [0]{\spacefactor3000\relax}%
\providecommand \BibitemShut  [1]{\csname bibitem#1\endcsname}%
\let\auto@bib@innerbib\@empty
\bibitem [{\citenamefont {Yoshioka}(2002)}]{2002-yoshioka-zl}%
  \BibitemOpen
  \bibfield  {author} {\bibinfo {author} {\bibfnamefont {D.}~\bibnamefont
  {Yoshioka}},\ }\href
  {http://www.loc.gov/catdir/enhancements/fy0813/2002021123-d.html} {\emph
  {\bibinfo {title} {The quantum Hall effect}}}\ (\bibinfo  {publisher}
  {Springer},\ \bibinfo {address} {Berlin},\ \bibinfo {year}
  {2002})\BibitemShut {NoStop}%
\bibitem [{\citenamefont {Hasan}\ and\ \citenamefont
  {Moore}(2011)}]{2011-hasan-uq}%
  \BibitemOpen
  \bibfield  {author} {\bibinfo {author} {\bibfnamefont {M.~Z.}\ \bibnamefont
  {Hasan}}\ and\ \bibinfo {author} {\bibfnamefont {J.~E.}\ \bibnamefont
  {Moore}},\ }\href {\doibase 10.1146/annurev-conmatphys-062910-140432}
  {\bibfield  {journal} {\bibinfo  {journal} {Annu. Rev. Cond. Mat. Phys.}\
  }\textbf {\bibinfo {volume} {2}},\ \bibinfo {pages} {55} (\bibinfo {year}
  {2011})}\BibitemShut {NoStop}%
\bibitem [{\citenamefont {Qi}\ and\ \citenamefont {Zhang}(2011)}]{2011-qi-uq}%
  \BibitemOpen
  \bibfield  {author} {\bibinfo {author} {\bibfnamefont {X.-L.}\ \bibnamefont
  {Qi}}\ and\ \bibinfo {author} {\bibfnamefont {S.-C.}\ \bibnamefont {Zhang}},\
  }\href {\doibase 10.1103/RevModPhys.83.1057} {\bibfield  {journal} {\bibinfo
  {journal} {Rev. Mod. Phys.}\ }\textbf {\bibinfo {volume} {83}},\ \bibinfo
  {pages} {1057} (\bibinfo {year} {2011})}\BibitemShut {NoStop}%
\bibitem [{\citenamefont {Balents}(2010)}]{2010-balents-uq}%
  \BibitemOpen
  \bibfield  {author} {\bibinfo {author} {\bibfnamefont {L.}~\bibnamefont
  {Balents}},\ }\href {\doibase 10.1038/nature08917} {\bibfield  {journal}
  {\bibinfo  {journal} {Nature}\ }\textbf {\bibinfo {volume} {464}},\ \bibinfo
  {pages} {199} (\bibinfo {year} {2010})}\BibitemShut {NoStop}%
\bibitem [{\citenamefont {Laughlin}(1981)}]{1981-laughlin-yq}%
  \BibitemOpen
  \bibfield  {author} {\bibinfo {author} {\bibfnamefont {R.~B.}\ \bibnamefont
  {Laughlin}},\ }\href {\doibase 10.1103/PhysRevB.23.5632} {\bibfield
  {journal} {\bibinfo  {journal} {Phys. Rev. B}\ }\textbf {\bibinfo {volume}
  {23}},\ \bibinfo {pages} {5632} (\bibinfo {year} {1981})}\BibitemShut
  {NoStop}%
\bibitem [{\citenamefont {Niu}\ \emph {et~al.}(1985)\citenamefont {Niu},
  \citenamefont {Thouless},\ and\ \citenamefont {Wu}}]{1985-niu-nr}%
  \BibitemOpen
  \bibfield  {author} {\bibinfo {author} {\bibfnamefont {Q.}~\bibnamefont
  {Niu}}, \bibinfo {author} {\bibfnamefont {D.~J.}\ \bibnamefont {Thouless}}, \
  and\ \bibinfo {author} {\bibfnamefont {Y.-S.}\ \bibnamefont {Wu}},\ }\href
  {\doibase 10.1103/PhysRevB.31.3372} {\bibfield  {journal} {\bibinfo
  {journal} {Phys. Rev. B}\ }\textbf {\bibinfo {volume} {31}},\ \bibinfo
  {pages} {3372} (\bibinfo {year} {1985})}\BibitemShut {NoStop}%
\bibitem [{\citenamefont {Wen}\ and\ \citenamefont {Zee}(1992)}]{1992-wen-fk}%
  \BibitemOpen
  \bibfield  {author} {\bibinfo {author} {\bibfnamefont {X.~G.}\ \bibnamefont
  {Wen}}\ and\ \bibinfo {author} {\bibfnamefont {A.}~\bibnamefont {Zee}},\
  }\href {\doibase 10.1103/PhysRevLett.69.953} {\bibfield  {journal} {\bibinfo
  {journal} {Phys. Rev. Lett.}\ }\textbf {\bibinfo {volume} {69}},\ \bibinfo
  {pages} {953} (\bibinfo {year} {1992})}\BibitemShut {NoStop}%
\bibitem [{\citenamefont {Iengo}\ and\ \citenamefont
  {Li}(1994)}]{1994-iengo-fk}%
  \BibitemOpen
  \bibfield  {author} {\bibinfo {author} {\bibfnamefont {R.}~\bibnamefont
  {Iengo}}\ and\ \bibinfo {author} {\bibfnamefont {D.}~\bibnamefont {Li}},\
  }\href {\doibase 10.1016/0550-3213(94)90010-8} {\bibfield  {journal}
  {\bibinfo  {journal} {Nuclear Physics B}\ }\textbf {\bibinfo {volume}
  {413}},\ \bibinfo {pages} {735} (\bibinfo {year} {1994})}\BibitemShut
  {NoStop}%
\bibitem [{\citenamefont {Read}(2009)}]{2009-read-fk}%
  \BibitemOpen
  \bibfield  {author} {\bibinfo {author} {\bibfnamefont {N.}~\bibnamefont
  {Read}},\ }\href {\doibase 10.1103/PhysRevB.79.045308} {\bibfield  {journal}
  {\bibinfo  {journal} {Phys. Rev. B}\ }\textbf {\bibinfo {volume} {79}},\
  \bibinfo {pages} {045308} (\bibinfo {year} {2009})}\BibitemShut {NoStop}%
\bibitem [{\citenamefont {Hoyos}\ and\ \citenamefont
  {Son}(2012)}]{2012-hoyos-fk}%
  \BibitemOpen
  \bibfield  {author} {\bibinfo {author} {\bibfnamefont {C.}~\bibnamefont
  {Hoyos}}\ and\ \bibinfo {author} {\bibfnamefont {D.~T.}\ \bibnamefont
  {Son}},\ }\href {\doibase 10.1103/PhysRevLett.108.066805} {\bibfield
  {journal} {\bibinfo  {journal} {Phys. Rev. Lett.}\ }\textbf {\bibinfo
  {volume} {108}},\ \bibinfo {pages} {066805} (\bibinfo {year}
  {2012})}\BibitemShut {NoStop}%
\bibitem [{\citenamefont {Bradlyn}\ \emph {et~al.}(2012)\citenamefont
  {Bradlyn}, \citenamefont {Goldstein},\ and\ \citenamefont
  {Read}}]{2012-bradlyn-fk}%
  \BibitemOpen
  \bibfield  {author} {\bibinfo {author} {\bibfnamefont {B.}~\bibnamefont
  {Bradlyn}}, \bibinfo {author} {\bibfnamefont {M.}~\bibnamefont {Goldstein}},
  \ and\ \bibinfo {author} {\bibfnamefont {N.}~\bibnamefont {Read}},\ }\href
  {\doibase 10.1103/PhysRevB.86.245309} {\bibfield  {journal} {\bibinfo
  {journal} {Phys. Rev. B}\ }\textbf {\bibinfo {volume} {86}},\ \bibinfo
  {pages} {245309} (\bibinfo {year} {2012})}\BibitemShut {NoStop}%
\bibitem [{\citenamefont {Biswas}(2013)}]{2013-biswas-wq}%
  \BibitemOpen
  \bibfield  {author} {\bibinfo {author} {\bibfnamefont {R.~R.}\ \bibnamefont
  {Biswas}},\ }\href@noop {} {\enquote {\bibinfo {title} {Semiclassical theory
  of viscosity in quantum hall states},}\ } (\bibinfo {year} {2013}),\ \Eprint
  {http://arxiv.org/abs/1311.7149} {arXiv:1311.7149} \BibitemShut {NoStop}%
\bibitem [{\citenamefont {Can}\ \emph {et~al.}(2014)\citenamefont {Can},
  \citenamefont {Laskin},\ and\ \citenamefont {Wiegmann}}]{2014-can-qy}%
  \BibitemOpen
  \bibfield  {author} {\bibinfo {author} {\bibfnamefont {T.}~\bibnamefont
  {Can}}, \bibinfo {author} {\bibfnamefont {M.}~\bibnamefont {Laskin}}, \ and\
  \bibinfo {author} {\bibfnamefont {P.}~\bibnamefont {Wiegmann}},\ }\href
  {\doibase 10.1103/PhysRevLett.113.046803} {\bibfield  {journal} {\bibinfo
  {journal} {Phys. Rev. Lett.}\ }\textbf {\bibinfo {volume} {113}},\ \bibinfo
  {pages} {046803} (\bibinfo {year} {2014})}\BibitemShut {NoStop}%
\bibitem [{\citenamefont {Lammert}\ and\ \citenamefont
  {Crespi}(2004)}]{2004-lammert-yu}%
  \BibitemOpen
  \bibfield  {author} {\bibinfo {author} {\bibfnamefont {P.}~\bibnamefont
  {Lammert}}\ and\ \bibinfo {author} {\bibfnamefont {V.}~\bibnamefont
  {Crespi}},\ }\href {\doibase 10.1103/PhysRevB.69.035406} {\bibfield
  {journal} {\bibinfo  {journal} {Phys. Rev. B}\ }\textbf {\bibinfo {volume}
  {69}},\ \bibinfo {pages} {035406} (\bibinfo {year} {2004})}\BibitemShut
  {NoStop}%
\bibitem [{\citenamefont {Bueno}\ \emph {et~al.}(2012)\citenamefont {Bueno},
  \citenamefont {Furtado},\ and\ \citenamefont
  {de~M.~Carvalho}}]{2012-bueno-rm}%
  \BibitemOpen
  \bibfield  {author} {\bibinfo {author} {\bibfnamefont {M.}~\bibnamefont
  {Bueno}}, \bibinfo {author} {\bibfnamefont {C.}~\bibnamefont {Furtado}}, \
  and\ \bibinfo {author} {\bibfnamefont {A.}~\bibnamefont {de~M.~Carvalho}},\
  }\href {\doibase 10.1140/epjb/e2011-20726-4} {\bibfield  {journal} {\bibinfo
  {journal} {Eur. Phys. J. B}\ }\textbf {\bibinfo {volume} {85}},\ \bibinfo
  {pages} {53} (\bibinfo {year} {2012})}\BibitemShut {NoStop}%
\bibitem [{\citenamefont {R\"uegg}\ and\ \citenamefont
  {Lin}(2013)}]{2013-ruegg-xy}%
  \BibitemOpen
  \bibfield  {author} {\bibinfo {author} {\bibfnamefont {A.}~\bibnamefont
  {R\"uegg}}\ and\ \bibinfo {author} {\bibfnamefont {C.}~\bibnamefont {Lin}},\
  }\href {\doibase 10.1103/PhysRevLett.110.046401} {\bibfield  {journal}
  {\bibinfo  {journal} {Phys. Rev. Lett.}\ }\textbf {\bibinfo {volume} {110}},\
  \bibinfo {pages} {046401} (\bibinfo {year} {2013})}\BibitemShut {NoStop}%
\bibitem [{\citenamefont {Merzbacher}(1998)}]{1998-merzbacher-qe}%
  \BibitemOpen
  \bibfield  {author} {\bibinfo {author} {\bibfnamefont {E.}~\bibnamefont
  {Merzbacher}},\ }\href@noop {} {\emph {\bibinfo {title} {Quantum
  mechanics}}},\ \bibinfo {edition} {3rd}\ ed.\ (\bibinfo  {publisher} {John
  Wiler \& Sons},\ \bibinfo {year} {1998})\BibitemShut {NoStop}%
\bibitem [{\citenamefont {Avishai}\ and\ \citenamefont
  {Luck}(2008)}]{2008-avishai-fk}%
  \BibitemOpen
  \bibfield  {author} {\bibinfo {author} {\bibfnamefont {Y.}~\bibnamefont
  {Avishai}}\ and\ \bibinfo {author} {\bibfnamefont {J.~M.}\ \bibnamefont
  {Luck}},\ }\href {\doibase 10.1088/1742-5468/2008/06/P06007} {\bibfield
  {journal} {\bibinfo  {journal} {J. Stat. Mech.}\ }\textbf {\bibinfo {volume}
  {2008}},\ \bibinfo {pages} {P06007} (\bibinfo {year} {2008})}\BibitemShut
  {NoStop}%
\bibitem [{\citenamefont {Haldane}(1983)}]{1983-haldane-rt}%
  \BibitemOpen
  \bibfield  {author} {\bibinfo {author} {\bibfnamefont {F.~D.~M.}\
  \bibnamefont {Haldane}},\ }\href {\doibase 10.1103/PhysRevLett.51.605}
  {\bibfield  {journal} {\bibinfo  {journal} {Phys. Rev. Lett.}\ }\textbf
  {\bibinfo {volume} {51}},\ \bibinfo {pages} {605} (\bibinfo {year}
  {1983})}\BibitemShut {NoStop}%
\end{thebibliography}
\end{document}